\documentclass[11pt,twoside]{article}
\usepackage{./asp2014}

\aspSuppressVolSlug
\resetcounters

\def\useaffilUW{{1}}
\def\useaffilLSST{{2}}
\def\useaffilSLAC{{3}}
\def\useaffilPrinceton{{4}}
\def\useaffilIPAC{{5}}
\def\useaffilBelgrade{{6}}
\def\useaffilNCSA{{7}}
\def\useaffilIN2P3{{8}}
\def\useaffilBNL{{9}}
\def\useaffilDavis{{10}}
\def\useaffilLPC{{11}}
\def\useaffilPenn{{12}}
\def\useaffilToronto{{13}}
\def\useaffilAmes{{14}}
\def\useaffilUA{{15}}
\def\useaffilNOAO{{16}}
\begin{document}

\title{The LSST Data Management System}
\author{
Mario~Juri\'c,$^{\useaffilUW}$
Jeffrey~Kantor,$^{\useaffilLSST}$
K-T~Lim,$^{\useaffilSLAC}$
Robert~H.~Lupton,$^{\useaffilPrinceton}$
Gregory~Dubois-Felsmann,$^{\useaffilIPAC}$
Tim~Jenness,$^{\useaffilLSST}$
Tim~S.~Axelrod,$^{\useaffilLSST}$
Jovan~Aleksi\'c,$^{\useaffilBelgrade}$
Roberta~A.~Allsman,$^{\useaffilLSST}$
Yusra~AlSayyad,$^{\useaffilUW}$
Jason~Alt,$^{\useaffilNCSA}$
Robert~Armstrong,$^{\useaffilPrinceton}$
Jim~Basney,$^{\useaffilNCSA}$
Andrew~C.~Becker,$^{\useaffilUW}$
Jacek~Becla,$^{\useaffilSLAC}$
Steven~J.~Bickerton,$^{\useaffilPrinceton}$
Rahul~Biswas,$^{\useaffilUW}$
James~Bosch,$^{\useaffilPrinceton}$
Dominique~Boutigny,$^{\useaffilSLAC,\useaffilIN2P3}$
Matias~Carrasco~Kind,$^{\useaffilNCSA}$
David~R.~Ciardi,$^{\useaffilIPAC}$
Andrew~J.~Connolly,$^{\useaffilUW}$
Scott~F.~Daniel,$^{\useaffilUW}$
Gregory~E.~Daues,$^{\useaffilNCSA}$
Frossie~Economou,$^{\useaffilLSST}$
Hsin-Fang~Chiang,$^{\useaffilNCSA}$
Angelo~Fausti,$^{\useaffilLSST}$
Merlin~Fisher-Levine,$^{\useaffilBNL}$
D.~Michael~Freemon,$^{\useaffilNCSA}$
Perry~Gee,$^{\useaffilDavis}$,
Philippe~Gris,$^{\useaffilLPC}$
Fabio~Hernandez,$^{\useaffilIN2P3}$
Joshua~Hoblitt,$^{\useaffilLSST}$
\v{Z}eljko~Ivezi\'{c},$^{\useaffilUW}$
Fabrice~Jammes,$^{\useaffilLPC}$
Darko~Jevremovi\'c,$^{\useaffilBelgrade}$
R.~Lynne~Jones,$^{\useaffilUW}$
J.~Bryce~Kalmbach,$^{\useaffilUW}$
Vishal~P.~Kasliwal,$^{\useaffilPrinceton,\useaffilPenn}$
K.~Simon~Krughoff,$^{\useaffilUW}$
Dustin~Lang,$^{\useaffilPrinceton,\useaffilToronto}$
John~Lurie,$^{\useaffilUW}$
Nate~B.~Lust,$^{\useaffilPrinceton}$
Fergal~Mullally,$^{\useaffilPrinceton,\useaffilAmes}$
Lauren~A.~MacArthur,$^{\useaffilPrinceton}$
Peter~Melchior,$^{\useaffilPrinceton}$
Joachim~Moeyens,$^{\useaffilUW}$
David~L.~Nidever,$^{\useaffilLSST,\useaffilUA}$
Russell~Owen,$^{\useaffilUW}$
John~K~Parejko,$^{\useaffilUW}$
J.~Matt~Peterson,$^{\useaffilLSST}$
Donald~Petravick,$^{\useaffilNCSA}$
Stephen~R.~Pietrowicz,$^{\useaffilNCSA}$
Paul~A.~Price,$^{\useaffilPrinceton}$
David~J.~Reiss,$^{\useaffilUW}$
Richard~A.~Shaw,$^{\useaffilNOAO}$
Jonathan~Sick,$^{\useaffilLSST}$
Colin~T.~Slater,$^{\useaffilUW}$
Michael~A.~Strauss,$^{\useaffilPrinceton}$
Ian~S.~Sullivan,$^{\useaffilUW}$
John~D.~Swinbank,$^{\useaffilPrinceton}$
Schuyler~Van~Dyk,$^{\useaffilIPAC}$
Veljko~Vuj\v{c}i\'c,$^{\useaffilBelgrade}$
Alexander~Withers,$^{\useaffilNCSA}$
Peter~Yoachim$^{\useaffilUW}$
\affil{$^\useaffilUW$University of Washington, Seattle, WA, U.S.A; \email{mjuric@astro.washington.edu}}
\affil{$^\useaffilLSST$LSST Project Management Office, Tucson, AZ, U.S.A.}
\affil{$^\useaffilSLAC$SLAC National Laboratory, Menlo Park, CA, U.S.A.}
\affil{$^\useaffilPrinceton$Princeton University, Princeton, NJ, U.S.A.}
\affil{$^\useaffilIPAC$Infrared Processing and Analysis Center, California Institute of Technology, Pasadena, CA, U.S.A.}
\affil{$^\useaffilBelgrade$Astronomical Observatory, Belgrade, Serbia}
\affil{$^\useaffilNCSA$National Center for Supercomputing Applications, Urbana, IL, U.S.A.}
\affil{$^\useaffilIN2P3$IN2P3 Computing Center, CNRS, Lyon-Villeurbanne, France}
\affil{$^\useaffilBNL$Brookhaven National Laboratory, Upton, NY, U.S.A.}
\affil{$^\useaffilDavis$University of California, Davis, CA, U.S.A.}
\affil{$^\useaffilLPC$LPC, IN2P3, CNRS, Clermont-Ferrand, France}
\affil{$^\useaffilPenn$University of Pennsylvania, Philadelphia, PA, U.S.A.}
\affil{$^\useaffilToronto$University of Toronto, Toronto, ON, Canada}
\affil{$^\useaffilAmes$SETI/NASA Ames Research Center, Moffett Field, CA, U.S.A.}
\affil{$^\useaffilUA$University of Arizona, Tucson, AZ, U.S.A.}
\affil{$^\useaffilNOAO$NOAO, Tucson, AZ, U.S.A.}
}

% This section is for ADS Processing.  There must be one line per author.
\paperauthor{Mario~Juri\'c}{mjuric@astro.washington.edu}{0000-0003-1996-9252}{University of Washington}{Department of Astronomy}{Seattle}{WA}{98195}{U.S.A}
\paperauthor{Jeffrey~Kantor}{jkantor@lsst.org}{}{LSST}{Data Management}{Tucson}{AZ}{85719}{U.S.A.}
\paperauthor{K-T~Lim}{ktl@slac.stanford.edu}{}{SLAC}{}{Menlo Park}{CA}{94025}{U.S.A.}
\paperauthor{Tim~S.~Axelrod}{taxelrod@lsst.org}{}{LSST}{Data Management}{Tucson}{AZ}{85719}{U.S.A.}
\paperauthor{Tim~Jenness}{tjenness@lsst.org}{0000-0001-5982-167X}{LSST}{Data Management}{Tucson}{AZ}{85719}{U.S.A.}
\paperauthor{Frossie~Economou}{frossie@lsst.org}{0000-0002-8333-7615}{LSST}{Data Management}{Tucson}{AZ}{85719}{U.S.A.}
\paperauthor{David~L.~Nidever}{dnidever@lsst.org}{0000-0002-1793-3689}{LSST}{Data Management}{Tucson}{AZ}{85719}{U.S.A.}
\paperauthor{Angelo~Fausti}{afausti@lsst.org}{0000-0002-8095-305X}{LSST}{Data Management}{Tucson}{AZ}{85719}{U.S.A.}
\paperauthor{Jonathan~Sick}{jsick@lsst.org}{0000-0003-3001-676X}{LSST}{Data Management}{Tucson}{AZ}{85719}{U.S.A.}
\paperauthor{Joshua~Hoblitt}{jhoblitt@lsst.org}{}{LSST}{Data Management}{Tucson}{AZ}{85719}{U.S.A.}
\paperauthor{J.~Matt~Peterson}{jpeterson2@lsst.org}{}{LSST}{Data Management}{Tucson}{AZ}{85719}{U.S.A.}
\paperauthor{Zeljko~Ivezic}{ivezic@astro.washington.edu}{}{University of Washington}{Department of Astronomy}{Seattle}{WA}{98195}{U.S.A.}
\paperauthor{R.~Lynne~Jones}{ljones.uw@gmail.com}{0000-0001-5916-0031}{University of Washington}{Department of Astronomy}{Seattle}{WA}{98195}{U.S.A.}
\paperauthor{Peter~Yoachim}{yoachim@uw.edu}{0000-0003-2874-6464}{University of Washington}{Department of Astronomy}{Seattle}{WA}{98195}{U.S.A.}
\paperauthor{Darko~Jevremovic}{darko@aob.rs}{}{Astronomical Observatory}{}{Belgrade}{}{11060}{Serbia}
\paperauthor{Veljko~Vujcic}{}{}{Astronomical Observatory}{}{Belgrade}{}{11060}{Serbia}
\paperauthor{Jovan~Aleksic}{}{}{Astronomical Observatory}{}{Belgrade}{}{11060}{Serbia}
\paperauthor{Richard~A.~Shaw}{}{}{NOAO}{}{Tucson}{AZ}{85719}{U.S.A.}
\paperauthor{Robert~Armstrong}{rearmstr@gmail.com}{}{Princeton University}{Department of Astrophysical Sciences}{Princeton}{NJ}{08544}{U.S.A.}
\paperauthor{James~Bosch}{jbosch@astro.princeton.edu}{0000-0003-2759-5764}{Princeton University}{Department of Astrophysical Sciences}{Princeton}{NJ}{08544}{U.S.A.}
\paperauthor{Merlin~Fisher-Levine}{merlin.fisherlevine@gmail.com}{0000-0001-9440-8960}{Brookhaven National Laboratory}{Department of Physics}{Upton}{NY}{11793}{U.S.A.}
\paperauthor{Vishal~P.~Kasliwal}{vishal.kasliwal@gmail.com}{0000-0001-7970-0760}{University of Pennsylvania}{Department of Physics and Astronomy, Center for Particle Cosmology}{Philadelphia}{PA}{19104}{U.S.A.}
\paperauthor{Robert~H.~Lupton}{rhl@astro.princeton.edu}{0000-0003-1666-0962}{Princeton University}{Department of Astrophysical Sciences}{Princeton}{NJ}{08544}{U.S.A.}
\paperauthor{Nate~B.~Lust}{natelust@linux.com}{}{Princeton University}{Department of Astrophysical Sciences}{Princeton}{NJ}{08544}{U.S.A.}
\paperauthor{Lauren~A.~MacArthur}{lauren@astro.princeton.edu}{}{Princeton University}{Department of Astrophysical Sciences}{Princeton}{NJ}{08544}{U.S.A.}
\paperauthor{Peter~Melchior}{peter.m.melchior@gmail.com}{0000-0002-8873-5065}{Princeton University}{Department of Astrophysical Sciences}{Princeton}{NJ}{08544}{U.S.A.}
\paperauthor{Paul~A.~Price}{price@astro.princeton.edu}{}{Princeton University}{Department of Astrophysical Sciences}{Princeton}{NJ}{08544}{U.S.A.}
\paperauthor{Michael~A.~Strauss}{strauss@astro.princeton.edu}{0000-0002-0106-7755}{Princeton University}{Department of Astrophysical Sciences}{Princeton}{NJ}{08544}{U.S.A.}
\paperauthor{John~D.~Swinbank}{swinbank@princeton.edu}{0000-0001-9445-1846}{Princeton University}{Department of Astrophysical Sciences}{Princeton}{NJ}{08544}{U.S.A.}
\paperauthor{Jacek~Becla}{becla@slac.stanford.edu}{}{SLAC}{}{Menlo Park}{CA}{94025}{U.S.A.}
\paperauthor{Fabrice~Jammes}{}{}{IN2P3}{LPC}{Clermon-Ferrand}{}{}{France}
\paperauthor{Matias~Carrasco~Kind}{mcarras2@illinois.edu}{0000-0002-4802-3194}{NCSA}{University of Illinois at Urbana-Champaign}{Urbana}{IL}{61801}{U.S.A.}
\paperauthor{Hsin-Fang~Chiang}{hchiang2@illinois.edu}{0000-0002-1181-1621}{NCSA}{University of Illinois at Urbana-Champaign}{Urbana}{IL}{61801}{U.S.A.}
\paperauthor{Donald~Petravick}{petravic@illinois.edu}{}{NCSA}{University of Illinois at Urbana-Champaign}{Urbana}{IL}{61801}{U.S.A.}
\paperauthor{Jason~Alt}{jalt@illinois.edu}{}{NCSA}{University of Illinois at Urbana-Champaign}{Urbana}{IL}{61801}{U.S.A.}
\paperauthor{Gregory~E.~Daues}{daues@illinois.edu}{}{NCSA}{University of Illinois at Urbana-Champaign}{Urbana}{IL}{61801}{U.S.A.}
\paperauthor{Alexander~Withers}{alexw1@illinois.edu}{}{NCSA}{University of Illinois at Urbana-Champaign}{Urbana}{IL}{61801}{U.S.A.}
\paperauthor{Jim~Basney}{jbasney@illinois.edu}{}{NCSA}{University of Illinois at Urbana-Champaign}{Urbana}{IL}{61801}{U.S.A.}
\paperauthor{D.~Michael~Freemon}{}{}{NCSA}{University of Illinois at Urbana-Champaign}{Urbana}{IL}{61801}{U.S.A.}
\paperauthor{Stephen~R.Pietrowicz}{srp@illinois.edu}{}{NCSA}{University of Illinois at Urbana-Champaign}{Urbana}{IL}{61801}{U.S.A.}
\paperauthor{Yusra~AlSayyad}{yusra@u.washington.edu}{}{University of Washington}{Department of Astronomy}{Seattle}{WA}{98195}{USA}
\paperauthor{Andrew~C.~Becker}{becker@astro.washington.edu}{0000-0001-6661-3043}{University of Washington}{Department of Astronomy}{Seattle}{WA}{98195}{USA}
\paperauthor{Rahul~Biswas}{rbiswas@uw.edu}{0000-0002-5741-7195}{University of Washington}{Department of
Astronomy and eScience Institute}{Seattle}{WA}{98195}{USA}
\paperauthor{Andrew~J.~Connolly}{ajc@astro.washington.edu}{0000-0001-5576-8189}{University of Washington}{Department of Astronomy}{Seattle}{WA}{98195}{USA}
\paperauthor{Scott~F.~Daniel}{scottvalscott@gmail.com}{}{University of Washington}{Department of Astronomy}{Seattle}{WA}{98195}{USA}
\paperauthor{J.~Bryce~Kalmbach}{jbkalmbach@gmail.com}{0000-0002-6825-5283}{University of Washington}{Department of Astronomy}{Seattle}{WA}{98195}{USA}
\paperauthor{K.~Simon~Krughoff}{krughoff@uw.edu}{0000-0002-4410-7868}{University of Washington}{Department of Astronomy}{Seattle}{WA}{98195}{USA}
\paperauthor{John~Lurie}{lurie@u.washington.edu}{0000-0002-8114-0835}{University of Washington}{Department of Astronomy}{Seattle}{WA}{98195}{USA}
\paperauthor{Joachim~Moeyens}{moeyensj@uw.edu}{0000-0001-5820-3925}{University of Washington}{Department of Astronomy}{Seattle}{WA}{98195}{USA}
\paperauthor{Russell~Owen}{rowen@u.washington.edu}{}{University of Washington}{Department of Astronomy}{Seattle}{WA}{98195}{USA}
\paperauthor{John~K.~Parejko}{parejkoj@u.washington.edu}{}{University of Washington}{Department of Astronomy}{Seattle}{WA}{98195}{USA}
\paperauthor{David~J.~Reiss}{reiss@uw.edu}{}{University of Washington}{Department of Astronomy}{Seattle}{WA}{98195}{USA}
\paperauthor{Colin~T.~Slater}{ctslater@uw.edu}{0000-0002-0558-0521}{University of Washington}{Department of Astronomy}{Seattle}{WA}{98195}{USA}
\paperauthor{Ian~S.~Sullivan}{sullii@uw.edu}{}{University of Washington}{Department of Astronomy}{Seattle}{WA}{98195}{USA}
\paperauthor{Gregory~Dubois-Felsmann}{gpdf@ipac.caltech.edu}{}{California Institute of Technology}{Infrared Processing and Analysis Center}{Pasadena}{California}{91125}{USA}
\paperauthor{David~R.~Ciardi}{ciardi@ipac.caltech.edu}{}{California Institute of Technology}{Infrared Processing and Analysis Center}{Pasadena}{California}{91125}{USA}
\paperauthor{Schuyler~Van~Dyk}{vandyk@ipac.caltech.edu}{}{California Institute of Technology}{Infrared Processing and Analysis Center}{Pasadena}{California}{91125}{USA}
\paperauthor{Philippe Gris}{gris@clermont.in2p3.fr}{}{IN2P3}{LPC}{Clermon-Ferrand}{}{}{France}
\paperauthor{Roberta~A.~Allsman}{}{}{LSST}{Data Management}{Tucson}{AZ}{85719}{U.S.A.}
\paperauthor{Dominique~Boutigny}{boutigny@in2p3.fr}{}{IN2P3}{CC-IN2P3}{Villeurbanne}{}{}{France}
\paperauthor{Perry~Gee}{pgee2000@gmail.com}{}{University of California, Davis}{Department of Physics}{Davis}{California}{95616}{USA}
\paperauthor{Steven~J.~Bickerton}{}{}{Princeton University}{Department of Astrophysical Sciences}{Princeton}{NJ}{08544}{U.S.A.}
\paperauthor{Dustin~Lang}{}{}{University of Toronto}{}{Toronto}{ON}{M5S 3H4}{Canada}
\paperauthor{Fergal~Mullally}{fergal.mullally@nasa.gov}{}{NASA}{SETI/NASA Ames Research Center}{Moffett Field}{CA}{94035}{U.S.A.}

\clearpage

\begin{abstract}

The Large Synoptic Survey Telescope (LSST; \citealt{2008arXiv0805.2366I}) is a
large-aperture, wide-field, ground-based survey system that will image
the sky in six optical bands from 320 to 1050 nm, uniformly covering approximately
$18,000$~deg$^2$ of the sky over 800 times. The LSST is currently
under construction on Cerro Pach\'{o}n in Chile, and expected to enter operations
in 2022. Once operational, the LSST will explore a wide range of
astrophysical questions, from discovering ``killer'' asteroids
to examining the nature of Dark Energy.

The LSST will generate on average 15 TB of data per night,
and will require a comprehensive Data Management system to reduce the
raw data to scientifically useful catalogs and images with minimum human
intervention. These reductions will result in a real-time alert stream, and
eleven data releases over the 10-year duration of LSST operations.
To enable this processing, the LSST project is developing a new,
general-purpose, high-performance, scalable, well documented, open source
data processing software stack for O/IR surveys. Prototypes of this stack
are already capable of processing data from existing cameras (e.g., SDSS,
DECam, MegaCam), and form the basis of the Hyper-Suprime Cam (HSC) Survey
data reduction pipeline.

\end{abstract}

\vspace{-0.3in}
\section{ The Large Synoptic Survey Telescope }

The Large Synoptic Survey Telescope (LSST; \url{http://lsst.org}) will be an
automated astronomical survey system that will survey approximately
$10,000$~deg$^2$ of the sky every few nights in six optical bands from 320
to 1050\,nm \citep{2008arXiv0805.2366I, lsstSRD}. Over the planned 10-year baseline
survey, it will uniformly and repeatedly image about $18,000$~deg$^2$ of the sky
over $800$~times. These data will be used to explore
a wide range of astrophysical questions, ranging from discovering
``killer'' asteroids to examining the nature of Dark Energy
\citep[e.g., see][]{2009arXiv0912.0201L}.

The LSST survey system consists of a large-aperture, wide-field, ground-based telescope \citep{2014SPIE.9145E..1AG}
currently being constructed on the El Pe\~n\'{o}n peak of Cerro Pach\'{o}n in the Chilean
Andes, a 3.2 gigapixel camera \citep{2010SPIE.7735E..0JK}, and a data management
system, described here.

In this paper, we review the science goals and technical design of the
LSST Data Management system
(LSST DM; described previously in \citealt{2007ASPC..376....3K} and \citealt{2010SPIE.7740E..1NK}).
We begin by describing the overall architecture of the system in
Section~\ref{sec:dm}, followed by a discussion of the planned data
products in Section~\ref{sec:dp}. In Section~\ref{sec:dmstack}, we
discuss the software stack being developed for processing and serving
of LSST data. We  conclude by pointing out opportunities that this new codebase
represents both for LSST and the astronomical software community as
a whole (in Section~\ref{sec:summary}).

\section{ The LSST Data Management System }
\label{sec:dm}

The rapid cadence and scale of the LSST observing program will produce
approximately 15 TB per night of raw imaging data\footnote{For
  comparison, the volume of all imaging data collected over a decade
  and published in SDSS Data Release 7 \citep{2009ApJS..182..543A} is approximately 16 TB.}. The large data volume, the real-time aspects
(to be explained below), and the complexity of processing involved makes it impractical to defer the data reduction to the LSST end-users. Instead, the data collected by the LSST system will be automatically reduced to scientifically useful catalogs and images by the LSST Data Management system.
\\

The principal functions of the LSST Data Management system are to:
\begin{itemize}
\item Within 60 seconds of observation, process the incoming stream of exposures by archiving raw images, generating alerts to new sources or sources whose properties have changed significantly, and updating the relevant catalogs (``Level 1'' data products).
\item Periodically reprocess the accumulated survey data to provide a
  uniform photometric and astrometric calibration \citep[e.g.,][]{LSE-180}, measure the
  properties of all detected objects, and characterize objects based on their time-dependent behavior. The results of such a processing run form a {\em Data Release} (DR), which is a static, self-consistent data set suitable for use in performing scientific analyses of LSST data and publication of the results (the ``Level 2'' data products). All data releases will be archived for the entire operational life of the LSST
archive, with the two most recent releases available in a
queryable database.
\item Facilitate the creation of added-value (``Level 3'') data products, by providing suitable software,
  application programming interfaces (APIs),
and computing infrastructure at the LSST data access centers. Provide
  processing, storage, and network bandwidth to enable user
  analyses of the data without the need for petabyte-scale data
  transfers.
\item Make all LSST data available to the community through interfaces that utilize
community-accepted standards   to the maximum possible extent. 
\end{itemize}

Over the ten years of LSST survey operations and 11 data releases, this processing will result in a cumulative {\em processed} data size
approaching 500 petabytes (PB) for imaging, and over 50 PB for the
catalog databases. The final data release catalog database alone is expected
to be approximately 15 PB in size. A more detailed overview of the data products will be given in
Section~\ref{sec:dp}.
\\

\begin{figure}[!t]
%
% NOTE NOTE NOTE: The source of this figure is in DMsandwich.pptx.
% Edit that file and save it as PDF when an update is needed.
%
\begin{center}
\includegraphics[width=0.7\textwidth,clip]{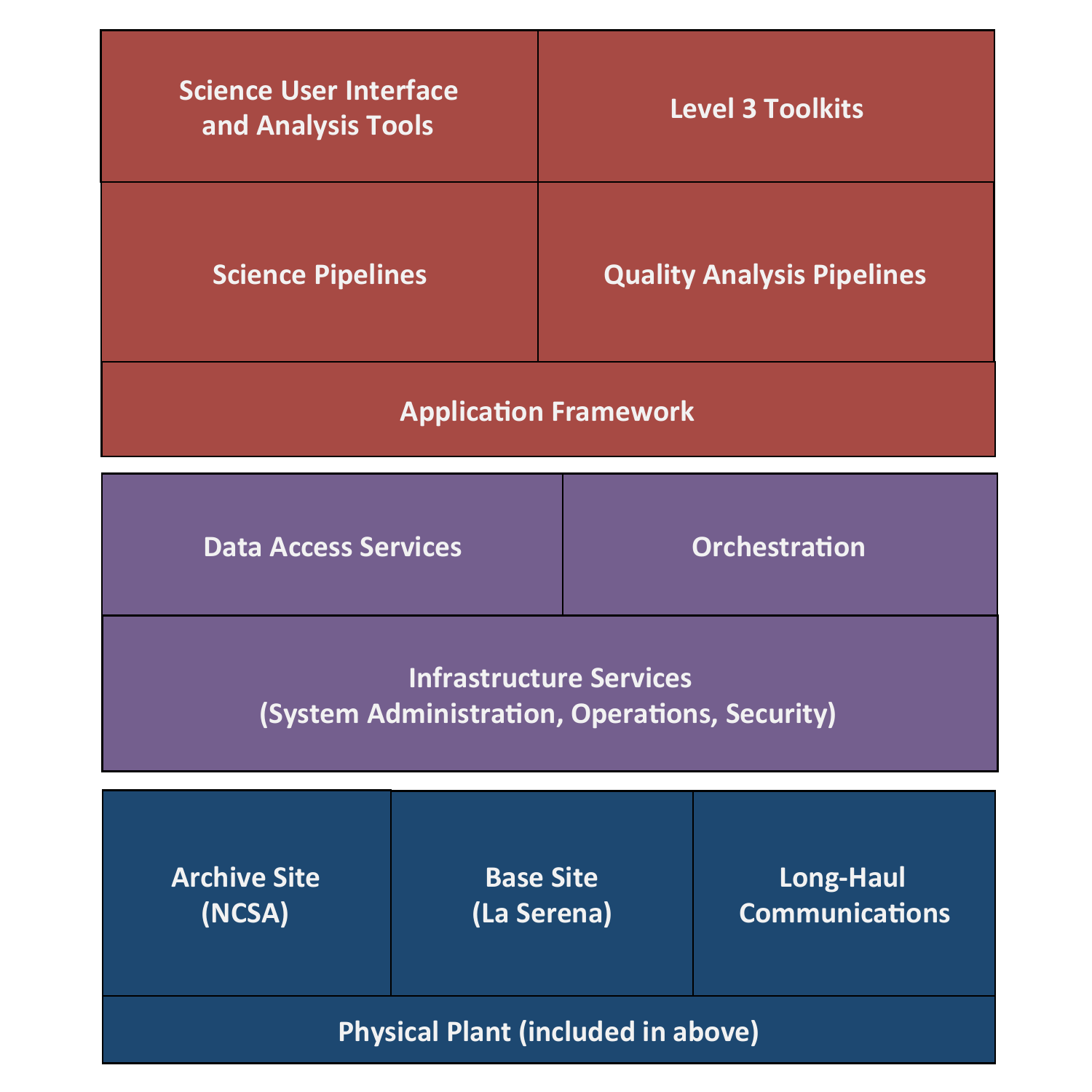}
\end{center}
\caption{The three-layered architecture of the data management system
(application [red, top], middleware [purple, middle], and infrastructure [blue, bottom] layers) enables scalability, reliability, and evolutionary capability.}
\label{Fig:DM1}
\end{figure}

The Data Management system is conceptually divided into three layers
\citep{LDM-148}: an infrastructure layer \citep{LDM-129} consisting of the computing, storage, and
networking hardware and system software; a middleware layer \citep{LDM-152}, which
handles distributed processing, data access, user interface, and
system operations services; and an applications layer
\citep{LDM-151,LDM-135,LDM-131}, which includes
the data pipelines and products and the science data archives (see
Fig.~\ref{Fig:DM1}).

\begin{figure*}[!t]
%
% NOTE NOTE NOTE: The source of this figure is in DMX2.pptx
% Edit that file and save it as PDF when an update is needed.
%
\begin{center}
\includegraphics[width=0.95\textwidth,clip]{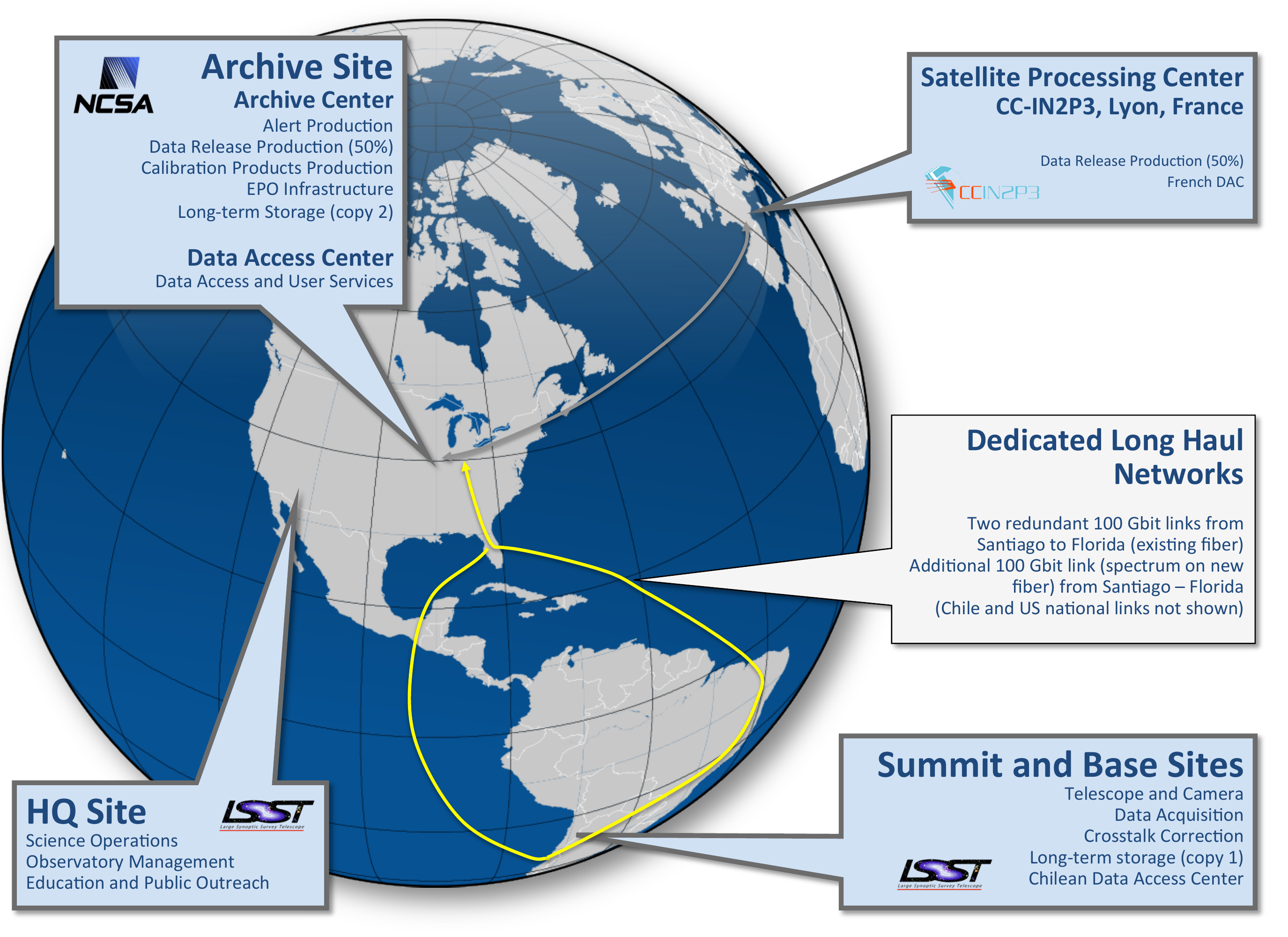}
\end{center}
\caption{The LSST data flow from the mountain summit/base facility in
Chile to the data access center and archive centers in the U.S. and France.
The data will be transported, in real time, over two (redundant) 
100 Gbit links from La Serena, Chile to Champaign, IL.
}
\label{Fig:DM2}
\end{figure*}

Physically, the DM system components will span four key facilities on three
continents: the Summit Facility at  Cerro Pach\'on (where the initial
detector cross-talk
correction will be performed), the Base Facility in La Serena (which will serve
as a retransmission hub for data
uploads to North America, as well as the data access center for the Chilean
community), the central Archive Facility at the National Center
for Supercomputing Applications (NCSA) in Champaign, Illinois, and a
satellite data processing center at Centre de Calcul de l'Institut National
de Physique Nucl\'{e}aire et de Physique des Particules (CC-IN2P3) in Lyon, France.
All Level 1 (nightly) and 50\% of Level 2 (data release) processing will take place at the
Archive Facility, which will also serve as a data access center
for the US community. The remaining 50\% of data processing will be performed at
the satellite center in Lyon.

The data will be transported between the centers over existing and new high-speed optical fiber
links from South America to the U.S., and from the U.S. to France (see Fig.~\ref{Fig:DM2}).
The data processing centers will have substantial computing
power (e.g., the central facility is expected to peak at $\sim1.6$~petaFLOPS of
compute power).
\\

\section{             LSST Data Products                    }
\label{sec:dp}

Data collected by the LSST camera and telescope will be automatically processed to {\em data products} -- catalogs, alerts,
and reduced images -- by the LSST Data Management system
(\S~\ref{sec:dm}). These products are designed to be sufficient to
enable a large majority of LSST science cases, without the need to
work directly with the raw pixels.  We give a high-level overview of
the LSST data products here; further details may be found in the LSST
Data Products Definition Document \citep{DPDD}.

\vskip 1em

Two major categories of data products will be produced and delivered by the DM
system:
\begin{itemize}
\item {\bf Level 1 data products}, designed to support the discovery,
  characterization, and rapid follow-up of time-dependent phenomena
  (``transient science''). These will be generated continuously every
  observing night, by detecting and characterizing sources in images
  differenced against deep templates. They will include alerts to
  objects that were newly discovered, or have changed brightness or
  position at a statistically significant level. The alerts to such
  events will be published within 60   seconds of observation
  \citep[see e.g.][for details]{2014htu..conf...19K}.\\
\\
In addition to transient science, Level 1 data products will support
discovery and follow-up of objects in the Solar System. Objects with
motions sufficient to cause trailing in a single exposure will be
identified and flagged as such when the alerts are broadcast. Those
that are not trailed will be identified and linked based on their
motion from observation to observation, over a period of a few
days. Their orbits will be published within 24 hours of
identification. The efficiency of linking (and thus the completeness
of the resulting orbit catalog) will depend on the final observing
cadence chosen for LSST, as well as the performance of the linking
algorithm \citep{LDM-156, 2015arXiv151103199J}.
\item {\bf Level 2 data products} are designed to enable systematics- and flux-limited science, and will be made available in annual Data Releases. These will include the (reduced and raw) single-epoch images, deep co-adds of the observed sky, catalogs of objects detected in LSST data, catalogs of sources (the detections and measurements of objects on individual visits), and catalogs of ``forced sources" (measurements of flux on individual visits at locations where objects were detected by LSST or other surveys). LSST data releases will also include fully reprocessed Level 1 data products, as well as all metadata and software necessary for the end-user to reproduce any aspect of LSST data release processing.\\
\\
A noteworthy aspect of LSST Level 2 processing is that it will largely
rely on {\bf multi-epoch model fitting}, or {\bf \em MultiFit}, to
perform near-optimal characterization of object properties. That is,
while the co-adds will be used to perform object {\em detection}, the
{\em measurement} of their properties will be performed by
simultaneously fitting (PSF-convolved) models to sets of single-epoch
observations. An extended source model -- a constrained linear
combination of two S\'ersic profiles -- and a point source model with
proper motion -- will generally be
fitted to each detected object\footnote{For performance reasons, it is
  likely that only the point source model will be fitted in the most
  crowded regions of the Galactic plane.}.\\
\\
Secondly, for the extended source model fits, the LSST will
characterize and store the shape of the associated likelihood surface
(and the posterior), and not just the maximum likelihood values and
covariances. The characterization will be accomplished by sampling,
with up to $\sim$200 independent likelihood samples retained for
each object. For storage cost reasons, these samples
may be retained only for those bands of greatest interest for
weak lensing studies.

\end{itemize}

The quality of all Level 1 and Level 2 data products will be extensively
assessed, both automatically as well as manually, using the Science
Data Quality Assurance (SDQA) pipelines \citep{LSE-63}.
\\

While a large majority of science cases will be adequately served by
Level 1 and 2 data products, a limited number of highly specialized
investigations may require custom, user-driven, processing of LSST
data. This processing will be most efficiently performed at the
LSST Archive Center, given the size of the LSST data set and the
associated storage and computational challenges. To enable such use
cases, the LSST DM system will devote the equivalent of 10\% of its
processing and storage capabilities to creation, use, and federation
of {\bf Level 3} (user-created) data products. It will also allow the
science teams to use the LSST database infrastructure to store and
share their results.

To further enable user-driven Level 3 processing, the {\em LSST Software
Stack}, described in Section~\ref{sec:dmstack}, has been explicitly
designed with reusability and extendability in mind, and will be made
available to the LSST user community. This will allow the LSST users to
more rapidly develop custom Level 3 processing codes, leveraging 15+
years of investment and experience put into LSST codes. In addition to
executing such customized codes at the LSST data centers, LSST users
will be able to run it on their own computational resources as well.\\

\section{The LSST Software Stack}
\label{sec:dmstack}

The LSST Software Stack\footnote{Source code and development versions are available at
\url{http://dm.lsst.org}.} aspires to be a well documented, state-of-the-art,
high-performance, scalable, multi-camera, open source, O/IR survey
data processing and analysis system, built to enable LSST survey data
reduction and the writing of custom, user-driven, code for Level 3
processing. Its design and implementation are led by a team distributed
across six institutions (the LSST Project Management Office, IPAC, NCSA,
Princeton University, SLAC National Laboratory and the University
of Washington) but also includes contributions from the broader LSST
community. Once completed, it will comprise of
all science pipelines \citep{LDM-151, LDM-156} needed to accomplish LSST data processing tasks
(e.g., calibration, single frame processing, co-addition, image
differencing, multi-epoch measurement, asteroid orbit determination,
etc.), quality assessment pipelines, the necessary data
access and orchestration middleware \citep{LDM-152}, the
image store and catalog database \citep{LDM-135}
as well as the science user interface components \citep[SUI;][]{LDM-131}.
The SUI components are planned to be based on IPAC's Firefly toolkit,
described by \citet{O10-1_adassxxv} elsewhere in this volume.
\\

Algorithm development for the LSST software builds on the expertise
and experience of prior large astronomical surveys such as
SDSS \citep{2000AJ....120.1579Y},
Pan-STARRS \citep{2006amos.confE..50M,2010SPIE.7733E..0EK},
DES \citep{DESDM},
SuperMACHO \citep{2005IAUS..225..357B},
ESSENCE \citep{2007ApJ...666..674M},
DLS \citep{2002SPIE.4836...73W},
CFHT-LS \citep{2012MNRAS.427..146H, 2013MNRAS.429.2858M, 2012AJ....143...38G},
and UKIDSS \citep{2007MNRAS.379.1599L}
to name a few, as well as software tools
\citep[e.g., SExtractor,][ascl:1010.064]{1996A&AS..117..393B}.
The pipelines written for these surveys have demonstrated that it is
possible to carry out largely autonomous data reduction of large data
sets, automated detection of sources and objects, and the
extraction of scientifically useful characteristics of those objects.

While firmly footed in this prior history, the LSST software stack has
largely been written anew, for reasons of performance, extendability, and
maintainability as well as stringent science and system quality requirements.
All LSST codes are being developed with the intent of
following software engineering best practices, including modularity, clear definition
of interfaces, continuous integration,
utilization of unit testing, a single set of documentation and coding
standards. The primary user-facing implementation language is Python and, where
necessary for performance reasons or when the algorithms require access to 
complex data structures, we use C++ with SWIG wrappers
\citep{Beazley:1996:SEU:1267498.1267513}.
See the contribution by \citet{P056_adassxxv} elsewhere in this volume for
more discussion on the architecture of the LSST software stack.
\\

\begin{figure}[!t]
%
% NOTE NOTE NOTE: The source of this figure is in DMStripe82.pptx
%
\includegraphics[width=0.9\hsize,clip]{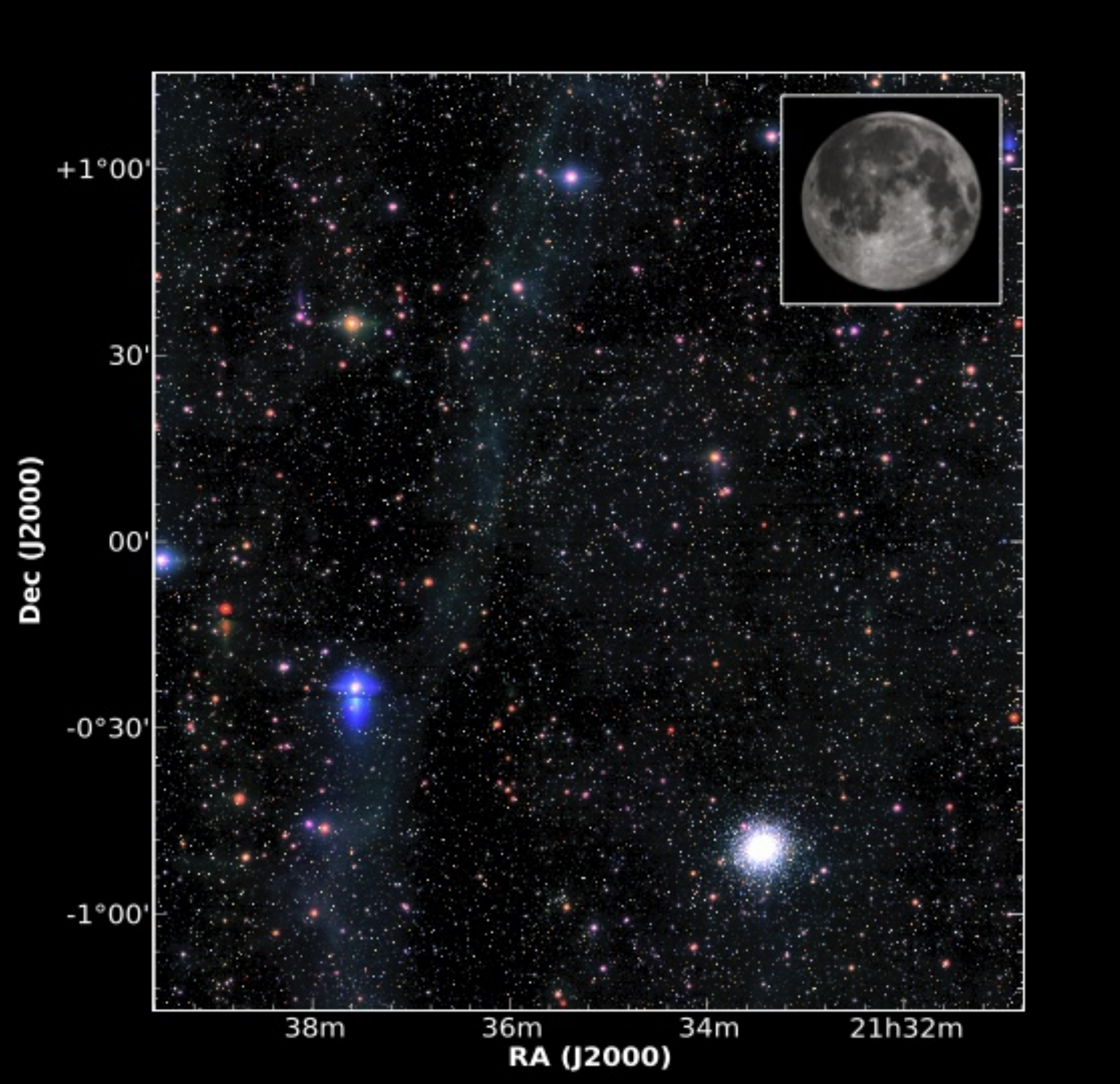}
\caption{
A $gri$ color composite of a small region in the vicinity of globular cluster M2, taken from a co-add of
SDSS Stripe 82 data produced with LSST software stack prototypes.  The
co-addition employs a novel ``background matching" technique
\citep{2014MNRAS.440.1296H} that improves
background estimation and preserves the diffuse structures in the resulting
co-add. The image of the Moon has been inserted for scale. The full
co-add can be browsed at \url{http://moe.astro.washington.edu/sdss}.}
\label{Fig:DMStripe82}
\end{figure}

The LSST data management software has been prototyped for over eight
years. It has been exercised in eight data challenges \citep[see e.g.][]{2010SPIE.7740E..1OK}, with increasing
degrees of complexity. Advanced prototypes of the distributed, shared-nothing, database 
being written for LSST ({\em ``Qserv''}) have been tested on
a 150-node cluster using 55 billion rows and 30 terabytes of
simulated data \citep{Wang:2011:QDS:2063348.2063364, qserv-s15tests}. Besides processing simulated LSST data
\citep{2014SPIE.9150E..14C, 0067-0049-218-1-14}, LSST science pipelines
have also been used to process images from CFHTLS
and SDSS \citep{2012ApJS..203...21A}. As an example,
Figure~\ref{Fig:DMStripe82} shows a small region in the vicinity of M2
taken from a large co-addition of SDSS Stripe 82 data, generated with LSST
software stack prototypes. Furthermore, the current development version
of the LSST software stack forms the basis for the data processing pipelines
of the Subaru Hyper-Suprime Cam Survey \citep{2012SPIE.8446E..0ZM} and
has been used successfully to produce two data releases of HSC Survey data.
\\

\section{               Summary              }
\label{sec:summary}

The LSST is envisioned to be a highly automated survey system
designed to enable investigations of phenomena on timescales ranging
from minutes to years, and covering approximately half of the sky
\citep{2009arXiv0912.0201L}. A comprehensive data management system,
such as the one described in this paper, is necessary to make possible
the research expected of such a facility.
\\

Though built to enable LSST science, we are hopeful that the individual
components of LSST data management system -- and especially the software
stack -- will be of broader utility.  The LSST software stack is free
software, licensed under the terms of the GNU General Public License,
Version 3.  The stack prototype code and documentation are available via
\url{http://dm.lsst.org}. Its open source nature, an open development process,
a long-term project commitment and a design that can be modified for use
with other cameras may make it useful for the processing of imaging data
beyond LSST.
\\

\acknowledgements

This material is based upon work supported in part by the National
Science Foundation through Cooperative Support Agreement (CSA) Award
No. AST-1227061 under Governing Cooperative Agreement 1258333 managed
by the Association of Universities for Research in Astronomy (AURA),
and the Department of Energy under Contract No. DE-AC02-76SF00515 with
the SLAC National Accelerator Laboratory.  Additional LSST funding
comes from private donations, grants to universities, and in-kind
support from LSSTC Institutional Members.

The authors would like to thank Steve Ritz, Sandrine Thomas, and Chuck
Claver for thorough reading of the manuscript and thoughtful comments
that improved it significantly.

\bibliography{O3-1}  % For BibTex

\end{document}